\documentclass[11pt,twoside]{article}
\usepackage{asp2004}
\usepackage{psfig}
\usepackage{epsf}
\usepackage{graphics}
\usepackage{lscape}
\usepackage{natbib}
\usepackage{amsmath}
\usepackage{xspace}

\def\edcomment#1{\iffalse\marginpar{\raggedright\sl#1\/}\else\relax\fi}
\reversemarginpar

\def\ie{{\it i.e.}}
\def\eg{{\it e.g.}}
\def\ltsima{$\; \buildrel < \over \sim \;$}
\def\simlt{\lower.5ex\hbox{\ltsima}}
\def\gtsima{$\; \buildrel > \over \sim \;$}
\def\simgt{\lower.5ex\hbox{\gtsima}}
\def\fesc{{$\langle f_{esc}\rangle$}\xspace}

\def\h2{H$_2$\xspace}
\def\ion#1#2{\text{#1\,\sc #2}}

\def\HII{{\ion{H}{ii} }}

\begin{document}
\title{Reionisation: the role of Globular Clusters}
\author{Massimo Ricotti}
\affil{Institute of Astronomy, Madingley Road, Cambridge CB3 0HA}

\begin{abstract}
  In this talk I discuss the role of proto-globular clusters as the
  dominant sources of radiation that reionised hydrogen in the
  intergalactic medium (IGM) at redshift $z \sim 6$.  Observations at
  lower redshift indicate that only a small fraction, \fesc, of
  hydrogen ionising radiation emitted from massive stars can escape
  unabsorbed by the galaxy into the IGM.  High redshift galaxies are
  expected to be more compact and gas rich than present day galaxies,
  consequently \fesc from their disks or spheroids might have been
  very small. But if the sites of star formation in the galaxies are
  off-centre and if the star formation efficiency of the
  proto-clusters is high, then the mean \fesc calculated for these
  objects only, is expected to be close to unity.
  
  Here I argue that this mode of star formation is consistent with
  several models for globular clusters formation.  Using simple
  arguments based on the observed number of globular cluster systems
  in the local universe and assuming that the oldest globular clusters
  formed before reionisation and had \fesc$\sim 1$, I show that they
  produced enough ionising photons to reionise the IGM at $z \sim 6$.
  I also emphasise that globular cluster formation might have been the
  dominant mode of star formation at redshifts from 6 to 12.
\end{abstract}

\section{Introduction}

In this talk, using simple arguments, I emphasise the important
cosmological role of globular cluster (GC) formation at high-redshift.
I show that the formation of GCs may have been the dominant mode of
star formation near the epoch of reionisation and have contributed
significantly to it. The material presented in this talk is based on
published work by \cite{RicottiS:00} and \cite{Ricotti:02}.

Observation of Ly$\alpha$ absorption systems toward high-redshift
quasars \citep{Becker:01} indicate that the redshift of reionisation
of the intergalactic medium (IGM) is $z_{\rm rei} \sim 6$. The recent
result from the WMAP satellite \citep{Kogut:03} of an early epoch of
reionisation will not be addressed in this talk. The reader can refer
to \cite{RicottiOI:03, RicottiO:03} if interested in this topic.

A key ingredient in determining the effectiveness by which galaxies
photoionise the surrounding IGM is the parameter \fesc, defined here
as the mean fraction of ionising photons escaping from galaxy halos
into the IGM.  Cosmological simulations and semi-analytical models of
IGM reionisation by stellar sources find that, in order to reionise
the IGM by $z =6 - 7$, the escape fraction from galaxies must be
relatively large: \fesc$\simgt 10$\% assuming a Salpeter initial mass
function (IMF) and the standard $\Lambda$CDM cosmological model.  The
assumption of a universal star formation efficiency (SFE) is
consistent with the observed values of the star formation rate (SFR)
at $0<z<5$ and total star budget at $z =0$. However, the assumption of
a constant \fesc does not seem to be consistent with observations.  An
escape fraction \fesc$\sim 1$ is required for reionisation at $z \sim
6$ but the ionising background at $z \sim 3$ is consistent with
\fesc$\simlt 10$\% \citep{Bianchi:01}.  Small values of \fesc at $z
\simlt 3$ are also supported by direct observations of the Lyc
emission from Lyman-break and starburst galaxies.  \cite{Giallongo:02}
find an upper limit \fesc$< 16$\% at $z \sim 3$ \citep[but
see][]{Steidel:01} and observations of low-redshift starbursts are
consistent with \fesc upper limits ranging from a few percent up to
10\% \citep{Hurwitz:97,Deharveng:01}.

Theoretical models \citep[\eg,][]{Dove:00} for the radiative
transfer of ionising radiation through the disk layer of spiral
galaxies similar to the Milky Way find \fesc$\sim 6$\%. At high
redshift the mean value of \fesc is expected to decrease almost
exponentially with increasing redshift \citep{RicottiS:00, Wood:00};
at $z > 6$, \fesc$\simlt 0.1-1$\% even assuming star formation rates
typical of starburst galaxies.  The majority of photons that escape
the halo come from the most luminous OB associations located in the
outermost parts of the galaxy.  Indeed \cite{RicottiS:00} have shown
that changing the luminosity function of the OB association and the
density distribution of the stars has major effects on \fesc (see
their Figs. 8 and 9).

A star formation mode, in which very luminous OB associations form in
the outer parts of galaxy halos, may explain the large \fesc
required for reionisation.  Globular clusters are possible
observable relics of such a star formation mode.  Their age
is compatible with formation at reionisation or earlier.  Because of
their large star density they survived tidal destruction and represent
the most luminous tail of the luminosity distribution of old OB
associations.  In \S~\ref{sec:mods} I show that several models for
the formation of proto-GCs imply an \fesc$\sim 1$.  I will also show
that the total amount of stars in GCs observed today is sufficient to
reionise the universe at $z \sim 6$ if their \fesc$\sim 1$. This
conclusion is reinforced if the GCs we observe today are only a
fraction, $1/f_{di}$, of primordial GCs as a consequence of mass
segregation and tidal stripping.

In \S~\ref{sec:rev} I briefly review a few observational properties
and in \S~\ref{sec:mods} formation theories of GCs that motivate the
assumption of \fesc$\sim 1$; in \S~\ref{sec:meth} I discuss the model
assumptions in light of GC observations and present the results.  In
\S~\ref{sec:disc} I present my conclusions.

\section{Condensed review on GC systems}\label{sec:rev}
  
Most galaxies have a bimodal GCs distribution indicating that luminous
galaxies experience at least two major episodes of GCs formation. The
bulk of the globulars in the main body of the Galactic halo appear to
have formed during a short-lived burst ($\sim 0.5-2$ Gyr) that took
place about 13 Gyrs ago. This was followed by a second burst
associated with the formation of the galactic bulges.

The method for determining the absolute age of GCs is based on fitting
the observed colour-magnitude diagram with theoretical evolutionary
tracks. The systematics in the evolutionary model and the
determination of the cluster distance are the major sources of errors.
Recent determinations of the absolute age of old GCs find $t_{gc} =
12.5 \pm 1.2$ Gyr \citep{Chaboyer:98}, consistent with radioactive
dating of a very metal-poor star in the halo of our galaxy
\citep{Cayrel:01}.  Relative ages of Galactic GCs can be determined
with greater accuracy, since many systematic errors can be eliminated.
In our Galaxy $\Delta t_{gc}=0.5$ Gyr, but differences in age between
GC systems in different galaxies could be $\Delta t_{gc} \sim 2$ Gyr
\citep{Stetson:96}.

The GC specific frequency is defined as the number, $N$, of GCs per
$M_V=-15$ of parent galaxy light, $S_N = N \times 10^{0.4(M_V+15)}$
(Harris \& van den Bergh 1981). The most striking characteristic is
that $S_N({\rm Ellipticals})>S_N({\rm Spirals})$.  $S_N = 0.5$ in
Sc/Ir galaxies (Harris, 1991), $S_N = 1$ in spirals of types Sa/Sb ,
$S_N =2.5$ in field ellipticals \citep{KunduII:01}.  Converting to
luminosity ($L_V/L_\odot=10^{-0.4(M_V-4.83)}$) we have $N =
(L/L_\odot)S_N/8.55 \times 10^7$. I can therefore calculate the
efficiency of GC formation defined as,
\begin{equation}
\epsilon_{gc}={M_{gc} \over M}={f_{di}Nm_{gc} \over M}={S_N f_{di} \over (M/L)_V}
\times 0.00585,
\label{eq:ef}
\end{equation}
where $M_{gc}$ is the total mass of the GC system, $m_{gc}=5 \times
10^5$ M$_\odot$ is the mean mass of GCs today, $M$ is the stellar
mass and $(M/L)_V$ is the mass to light ratio of the galaxy. In the
next paragraph we show that, because of dynamical evolution, $m_{gc}$
and $N$ are expected to be larger at the time of GC formation than
today.  Therefore, the parameter $f_{di} \ge 1$ is introduced to
account for dynamical disruption of GCs during their lifetime.

The IMF of GCs is not known.  The present mass function is known only
between 0.2 and 0.8 M$_\odot$ since high-mass stars are lost because of
two-body relaxation and stellar evolution processes.
Theoretical models show that the shape is consistent with a Salpeter-like
IMF.  The mean metallicity of old GCs is $Z \sim 0.03~Z_\odot$. One of
the most remarkable properties of GCs is the uniformity of their
internal metallicity $\Delta[Fe/H] \simlt 0.1$. This implies that the
bulk of the stars that constitute a GC formed in a single monolithic
burst of star formation. A typical GC emits $S \approx 3 \times
10^{53}$ s$^{-1}$ ionising photons in a burst lasting 4 Myrs: about
$300$ times the ionising luminosity of largest OB associations in our
Galaxy.

During their lifetime GCs lose a large part of their initial mass or
are completely destroyed by internal and external processes.
Numerical simulations show that about 50\%--90\% of the mass of GCs is
lost due to external processes, depending on the host galaxy
environment, initial concentration and IMF of the proto-GCs
\citep{Chernoff:90, GnedinO:97}. Many of the low-metallicity halo
field stars in the Milky-Way could be debris of disrupted GCs. The
mass in stars in the halo is about 100 times the mass in GCs.
Therefore the parameter $f_{di}$, defined in \S~\ref{sec:rev}, could
be as large as $f_{di}=100$. Overall $f_{di}$ is not well constrained
since it depends on unknown properties of the proto-GCs. According to
results of N-body simulations $f_{di}$ should be in the $f_{di} \sim
2-10$ range.

\section{Why is \fesc$\sim 1$ plausible for GCs?}\label{sec:mods}

I discuss separately two issues: (i) the
\fesc from the gas cloud in which the GC forms, and (ii) the \fesc
through any surrounding gas in the galaxy.
\begin{itemize}
\item[\bf{(i)}] The evidence for \fesc$\sim 1$ comes from the observed
  properties of present-day GCs. The fact that they are compact
  self-gravitating systems with low and uniform metallicity points to
  a high efficiency of conversion of gas into stars. A longer
  timescale of star formation would have enriched the gas of metals
  and the mechanical feedback from SN explosions would have stopped
  further star formation. If $f_*\approx 10$\% of the gas is converted
  into stars in a single burst (with duration $< 4$ Myr) at the centre
  of a spheroidal galaxy, following the simple calculations shown in
  \cite{RicottiS:00} [see their eq.~(18)] at $z=6$ we have,
  $f_{esc}=1-0.06(1-f_*)^2/f_* \sim 50\%$.
\item[{\bf(ii)}] The justification for \fesc$\sim 1$ is model
  dependent but in general there are two main arguments: a) the high
  efficiency of star formation $f_*$, and b) the sites of proto-GC
  formation in the outermost parts of the galaxy halo.
  
  In the {\it ``cosmological objects model''} ($30<z_f<7$) of
  \cite{Peebles:68} GCs form with efficiency $f_* \approx 100\%$,
  implying \fesc$=1$ (note that such a high $f_*$ is not found in
  numerical simulations of first object formation
  \citep{RicottiGSa:02, RicottiGSb:02}). In {\it ``hierarchical
    formation models''} ($10<z_f<3$) \citep[\eg,]{Harris:94,
    McLaughlin:96} GCs form in the disk or spheroid of galaxies with
  gas mass $M_g \sim 10^7-10^9$ M$_\odot$. Compact GCs survive the
  accretion by larger galaxies while the rest of the galaxy is tidally
  stripped. Assuming that $1-10$ GCs form in a galaxy with $M_g \sim
  10^7-10^8$ M$_\odot$ implies $f_* \sim 10\%$ and therefore \fesc$\ge
  50\%$. \fesc is larger if proto-GCs are located off-centre (\eg, if
  they form from cloud-cloud collisions during the galaxy assembly) or
  if part of the gas in the halo is collisionally ionised as a
  consequence of the virialization process.\\ In models such as the
  {\it ``super-shell fragmentation''} ($z_f<10$) of
  \cite{Taniguchi:99} or the {\it ``thermal instability''} ($z_f<7$)
  of \cite{Fall:85}, \fesc$\approx 1$ since proto-GCs form in the
  outermost part of an already collisionally ionised halo.
\end{itemize}
In summary, since \fesc depends strongly on the luminosity of the OB
associations and on their location, proto-GCs, being several hundred
times more luminous than Galactic OB associations, should have a
comparably larger \fesc.

\section{Method and Results}\label{sec:meth}

In this section I estimate the number of ionising photons emitted per
baryon per Hubble time, ${\cal N}_{ph}$, by GC formation. In
\S~\ref{ssec:simp} I derive ${\cal N}_{ph}$ assuming that all GCs
observed at $z=0$ formed in a time period $\Delta t_{gc}$ with
constant formation rate.  In \S~\ref{ssec:ps} I use the
Press-Schechter formalism to model more realistically the formation
rate of old GCs.
\begin{table}
\centering
\caption{Star census at z=0.}\label{tab:one}
\bigskip
\begin{tabular}[]{l|c|c|c|c}
\tableline
Type & $\omega_*$ (\%) & $S_N$ & (M/L)$_V$ & $\epsilon_{gc}$ (\%)\\
\tableline
Sph & $6.5_{-3}^{+4}$ & $2.4 \pm 0.4$ & $5.4 \pm 0.3$ & $0.26\pm 0.06$\\
Disk & $2_{-0.5}^{+1.5}$ & $1 \pm 0.1$ & $1.82 \pm 0.4$ & $0.32\pm 0.1$ \\
Irr & $0.15_{-0.05}^{+0.15}$ & 0.5 & $1.33 \pm 0.25$ & $0.22\pm 0.04$\\
Total & $9_{-3.5}^{+5.5}$ & - & - &  $0.3\pm 0.07$\\
\tableline
\tableline
\end{tabular}
\end{table}

\subsection{The simplest estimate}\label{ssec:simp}

I start by estimating the fraction, $\omega_{gc}$, of cosmic baryons
converted into GC stars. By definition
$\omega_{gc}=\omega_*\epsilon_{gc}$, where $\omega_*$ is the fraction
of baryons in stars at $z=0$ and $\epsilon_{gc}$ is the efficiency of
GC formation defined in \S~\ref{sec:rev}. In all the calculations I
assume $\Omega_b=0.04$.  In Table~\ref{tab:one}, I summarise the star
census at $z=0$ according to \cite{Persic:92} and I derive
$\epsilon_{gc}$ using eq.~(\ref{eq:ef}), assuming $f_{di}=1$. Using
similar arguments \cite{McLaughlin:99} finds a universal efficiency of
globular cluster formation $\epsilon_{gc}=(0.26 \pm 0.05)$\%, in
agreement with the simpler estimate presented here.  It follows that
$\omega_{gc}=f_{di}(2.7^{+2.3}_{-1.7} \times 10^{-4})$ at $z=0$.

The total number of ionising photons per unit time emitted by GCs is
$\eta\omega^f_{gc}/\Delta t_{gc}$, where $\eta$ is the number of
ionising photons emitted per baryon converted into stars, and
$\omega^f_{gc} \approx 2.1 \omega_{gc}$ takes into account the mass
loss due to stellar winds and SN explosions adopting an
instantaneous-burst star formation law. GCs did not recycle this lost
mass since they formed in a single burst of star formation.  $\eta$
depends on the IMF and on the metallicity of the star. I calculate
$\eta$ using a Salpeter IMF and star metallicity $Z =0.03~Z_\odot$
(see \S~\ref{sec:rev}) with Starburst99 code \citep{Leitherer:99},
and find $\eta=8967$.  The number of ionising photons per baryon
emitted in a Hubble time at $z=6$ is,
\begin{equation}
{\cal N}^{gc}_{ph}=\eta \omega^f_{gc} {t_H(z=6) \over \Delta 
t_{gc}}=(5.1^{+4.3}_{-3.2}){f_{di} \over \Delta t_{gc}~(Gyr)},
\label{eq:ns}
\end{equation}
where I have assumed \fesc$=1$ and Hubble time at $z=6$ $t_H=1\pm 0.1$
Gyr. I expect $1 \le f_{di} \simlt 100$ and $0.5 \simlt \Delta t_{gc}
\simlt 2$ Gyr. A conservative estimate of $f_{di} \simgt 2$ and
$\Delta t_{gc} \simlt 2$ Gyr (\ie, $10 < z_f < 3$) implies
$f_{di}/\Delta t_{gc} \simgt 1$. The IGM is reionised when ${\cal
N}_{ph}=C_\HII$, where $C_\HII=\langle n_{HII}^2 \rangle/\langle n_{HII}
\rangle^2$ is the ionised IGM clumping factor. According to the
adopted definition of \fesc, $C_\HII=1$ for a homogeneous IGM, or $1 \simlt
C_\HII \simlt 10$ taking into account IGM density fluctuations producing
the Ly$\alpha$ forest \citep{Miralda:00, Gnedin:00}. The estimate from
eq.~(\ref{eq:ns}) is rather rough because I have implicitly assumed
that the SFR is constant during the period of GC formation $\Delta
t_{gc}$. A more realistic SFR as a function of redshift requires
assuming a specific model for the formation of GCs.  I try to address
this question in the next section.

\subsection{Using the Press-Schechter formalism}\label{ssec:ps}
\begin{figure}[htb]
\plottwo{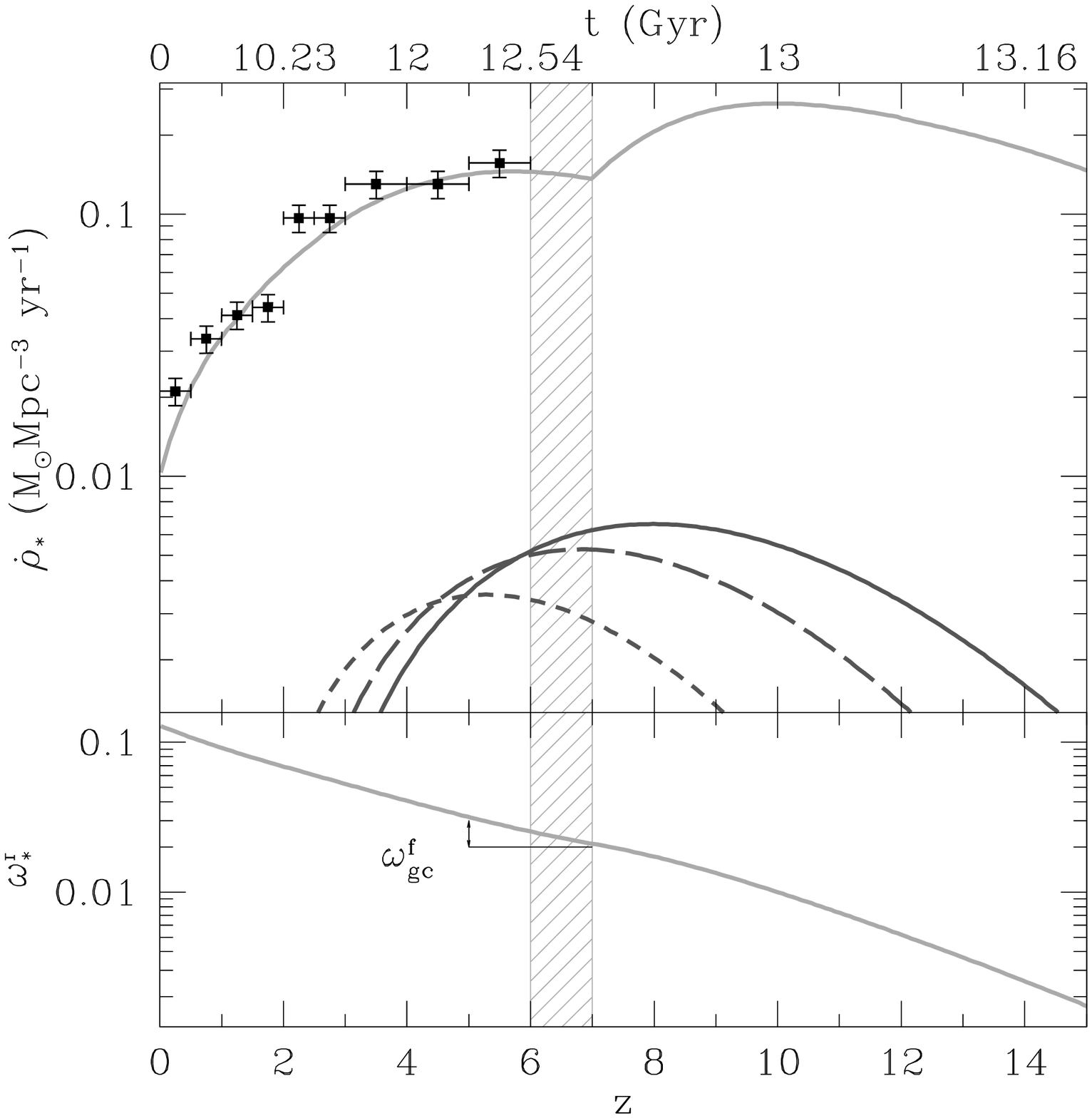}{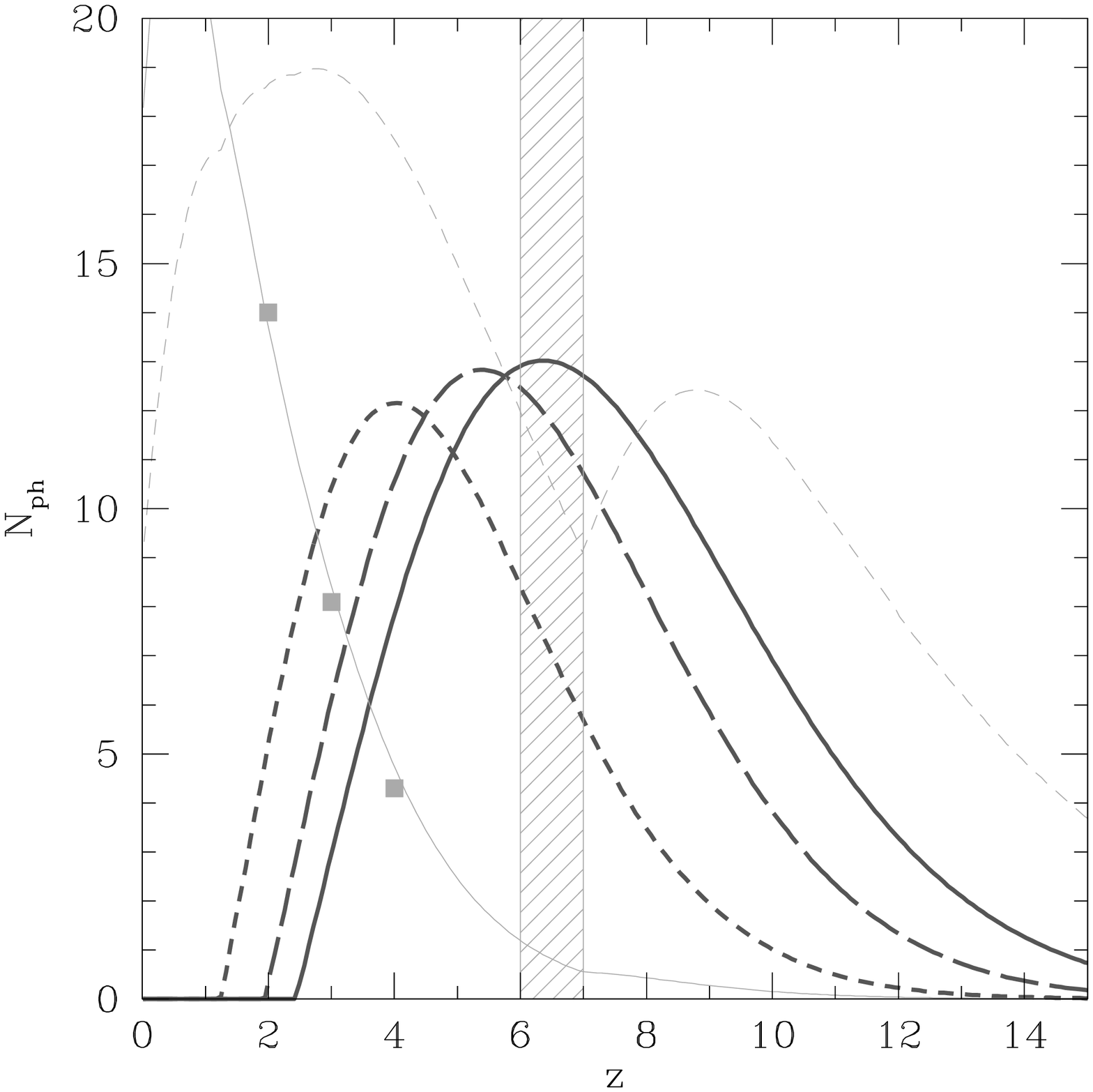}
\caption{\label{fig:sfr} ({\em left}) The thin solid line in the top panel shows the
  comoving SFR of galaxies as a function of time in our model. The
  thick solid, dashed and short-dashed lines show the SFR of GCs for
  cases (i), (ii) and (iii), respectively (assuming $f_{di}=2$). The
  bottom panel shows the stellar mass budget (in units of the baryon
  abundance), $\omega^f_*$, as a function of time. The segment with
  arrows is a visual aid to compare the GC contribution,
  $\omega^f_{gc}$, (assuming $f_{di}=20$) to $\omega^f_*$ around $z
  \sim 6$. ({\em right}) Emissivity (photons per baryon per Hubble
  time) as a function of redshift. The thick solid, dashed and
  short-dashed lines show the contribution of GCs for cases (i), (ii)
  and (iii), respectively (assuming $f_{di}=2$).  The thin lines show
  the contribution of galaxies assuming a realistic \fesc$=0.1 \times
  \exp[-z/2]$ (solid) and a constant \fesc$=5$\% (dashed). [Plots from
  Ricotti (2002).]}
\end{figure}

I assume that the formation rate of stars or GCs in galaxies is
proportional to the merger rate of galaxy halos (each galaxy undergoes
a major star burst episode when it virializes). Using the
Press-Schechter formalism I calculate,
\begin{eqnarray}
{d \omega^f_{gc}(z) \over dt}&=& {\cal A} \int_{M_1}^{M_2} {d \Omega(M_{dm}, z) \over
  dt} d\ln M_{dm},\label{eq:om1}\\
{d \omega^f_*(z) \over dt}&=& {\cal B} \int_{M_m}^\infty {d \Omega(M_{dm}, z) \over
  dt} d\ln M_{dm},
\label{eq:om}
\end{eqnarray}
where $\Omega(M_{dm},z)d\ln(M_{dm})$ is the mass fraction in
virialized dark matter halos of mass $M_{dm}$ at redshift z. I determine the
constants ${\cal A}$ and ${\cal B}$\footnote{By definition
$\int_0^\infty \Omega(M_{dm},z)d\ln(M_{dm})=1$. I find the following
values of the constants ${\cal A}=1.3\%, 1.6\%,0.6\%$ for cases (i),
(ii) and (iii) respectively (see text) and ${\cal B}=12\%$.} by
integrating eqs.~(\ref{eq:om1})-(\ref{eq:om}) with respect to time,
and assuming $\omega^f_{gc} = 0.1$\% (\ie, $f_{di}=2$) and
$\omega^f_*=1.4\omega_*=13$\% at $z=0$ (the factor 1.4 takes into
account the mass loss due to stellar winds and SN explosions adopting
a continuous star formation law).  I assume that GCs form in halos
with masses $M_1<M_{dm}<M_2$. The choice of $M_1$ and $M_2$ determine
the mean redshift, $z_f$, and time period, $\Delta t_{gc}$, for the
formation of old GCs. In order to be consistent with observations I
consider three cases: case (i) halos with virial temperature $2 \times
10^4 < T_{vir} < 5 \times 10^4$ K; case (ii) $5 \times 10^4 < T_{vir}
< 10^5$ K; and case (iii) $10^5 < T_{vir} < 5 \times 10^5$ K. In case
(i),(ii) and (iii) $\Delta t_{gc}=2.2, 3.7$ and 5.2 Gyr, respectively,
and the GC formation rate has a peak at $z = 7.5, 6$ and 4.6,
respectively. Disk and spheroid stars form in halos with
$M_{dm}>M_m$. At $z >10$ I assume that the first objects form in halos
with $M_m$ corresponding to a halo virial temperature $T_{vir}=5
\times 10^3$ K. At $z <10$ only objects with $T_{vir} > 2 \times 10^4$
K can form \citep[see][]{RicottiGSb:02}. The comoving star formation
rate, given by $\dot \rho_* = \overline \rho \dot \omega^f_*$, where
$\overline \rho = 5.51 \times 10^9$ M$_\odot$ Mpc$^{-3}$ is the mean
baryon density at $z=0$, is shown in Fig.~\ref{fig:sfr}(left). The points
show the observed SFR from \cite{Lanzetta:02}.

In Fig.~\ref{fig:sfr}(right) I show ${\cal N}_{ph}$ for GCs (thick
lines) and for galaxies (thin lines) defined as,
\begin{eqnarray}
{\cal N}^{gc}_{ph}&=& \eta f_{di}{d \omega^f_{gc} \over dt} t_{H}(z),\\
{\cal N}^*_{ph}&=& \eta \langle f_{esc}\rangle {d \omega^f_* \over dt}
t_{H}(z).
\end{eqnarray}
The thick solid, dashed and short-dashed lines show ${\cal
  N}^{gc}_{ph}$ for case (i), (ii) and (iii), respectively. For
comparison, I show (thin solid line) ${\cal N}^*_{ph}$ assuming
\fesc$=0.1 \times \exp[-z/2]$, derived theoretically by
\cite{RicottiS:00} and normalised to fit the observed values (squares)
of ${\cal N}^*_{ph}$ at $z=2,3,4$ \citep{Miralda:00}. The thin dashed
line shows ${\cal N}^*_{ph}$ assuming constant \fesc$=5$\%.

\section{Conclusions}\label{sec:disc}

The observed Lyman break galaxies at $z \sim 3$ are probably the most
luminous starburst galaxies of a population that produced the bulk of
the stars in our universe. Their formation epoch corresponds to the
assembly of the bulges of spirals and ellipticals. Nevertheless the
observed upper limit on \fesc from Lyman break galaxies, \fesc$\simlt
10$\%, may be insufficient to reionise the IGM according to numerical
simulations.  \cite{Ferguson:02}, using different arguments based on
the presence of an older stellar population, also noticed that the
radiation emitted from Lyman break galaxies at $z>3$ was insufficient
to reionise the IGM assuming a continuous star formation mode.

I propose that GCs during their formation may have produced enough
ionising photons to reionise the IGM.  Assuming $f_{di}=2$ (\ie,
during their evolution GCs have lost half of their original mass), I
find a stellar mass fraction in GCs, $\omega^f_{gc} \approx 0.1$\%,
small compared to the total stellar budget $\omega^f_* \sim 10$\% at
$z=0$.  But GCs are about 12-13 Gyr old and, if they formed between
$5<z<7$ (in about 0.5 Gyr), the expected total $\omega^f_*$ formed
during this time period is about $\omega^f_* \sim 1$\%, only 10 times
larger than $\omega^f_{gc}$. Assuming $f_{di}=20$, expected from the
results of N-body simulations, I find $\omega^f_{gc} \approx 1$\%,
suggesting that GC formation is an important mode of star formation at
high-redshift.  The star formation mode required to explain the
formation of GCs suggests an \fesc$\sim 1$ from these objects.  This
is because the mean \fesc is dominated by the most luminous OB
associations and GCs are extremely luminous, emitting $S \sim 3 \times
10^{53}$ s$^{-1}$ ionising photons in bursts lasting only 4 Myrs.
Moreover, according to many models, GCs form in the hot,
collisionally-ionised galaxy halo, from which all the ionising
radiation emitted can escape into the IGM.  Therefore it is not too
surprising, if GCs started forming before $z = 6$, that their expected
contribution to reionisation is significant. I find that the number of
ionising photons per baryon emitted in a Hubble time at $z=6$ by GCs
is ${\cal N}^{gc}_{ph}=(5.1^{+4.3}_{-3.2})f_{di}/\Delta t_{gc}>1$,
therefore sufficient to reionise the IGM even if we assume $f_{di}=1$.
Here, $\Delta t_{gc} \sim 0.5-2$ is the period of formation of the
bulk of old GCs in Gyrs. Using simple calculations based on
Press-Schechter formalism [see Fig.~\ref{fig:sfr}(right)] I find that,
if galaxies have \fesc$ \simlt 5$\%, GC contribution to reionisation
is important. If GCs formed by thermal instability in the halo of
$T_{vir} \sim 10^5$ K galaxies (case (iii)), the ionising sources have
a large bias (\ie, they form in rare peaks of the initial density
field). Therefore, the mean size of intergalactic \HII regions before
overlap is large and reionisation is inhomogeneous on large scales.

\bibliographystyle{/home/ricotti/Latex/TeX/apj}
\bibliography{/home/ricotti/Latex/TeX/archive}

\begin{thebibliography}{}

\bibitem[\protect\citeauthoryear{{Becker} et~al.}{{Becker}
  et~al.}{2001}]{Becker:01}
{Becker}, R.~H., et~al. 2001, \aj, 122, 2850

\bibitem[\protect\citeauthoryear{{Bianchi}, {Cristiani}, \& {Kim}}{{Bianchi}
  et~al.}{2001}]{Bianchi:01}
{Bianchi}, S., {Cristiani}, S.,  \& {Kim}, T.-S. 2001, \aap, 376, 1

\bibitem[\protect\citeauthoryear{{Cayrel} et~al.}{{Cayrel}
  et~al.}{2001}]{Cayrel:01}
{Cayrel}, R., et~al. 2001, Nature, 409, 691

\bibitem[\protect\citeauthoryear{{Chaboyer} et~al.}{{Chaboyer}
  et~al.}{1998}]{Chaboyer:98}
{Chaboyer}, B., {Demarque}, P., {Kernan}, P.~J.,  \& {Krauss}, L.~M. 1998,
  \apj, 494, 96

\bibitem[\protect\citeauthoryear{{Chernoff} \& {Weinberg}}{{Chernoff} \&
  {Weinberg}}{1990}]{Chernoff:90}
{Chernoff}, D.~F.,  \& {Weinberg}, M.~D. 1990, \apj, 351, 121

\bibitem[\protect\citeauthoryear{{Deharveng} et~al.}{{Deharveng}
  et~al.}{2001}]{Deharveng:01}
{Deharveng}, J.-M., {Buat}, V., {Le Brun}, V., {Milliard}, B., {Kunth}, D.,
  {Shull}, J.~M.,  \& {Gry}, C. 2001, \aap, 375, 805

\bibitem[\protect\citeauthoryear{{Dove}, {Shull}, \& {Ferrara}}{{Dove}
  et~al.}{2000}]{Dove:00}
{Dove}, J.~B., {Shull}, J.~M.,  \& {Ferrara}, A. 2000, \apj, 531, 846

\bibitem[\protect\citeauthoryear{{Fall} \& {Rees}}{{Fall} \&
  {Rees}}{1985}]{Fall:85}
{Fall}, S.~M.,  \& {Rees}, M.~J. 1985, \apj, 298, 18

\bibitem[\protect\citeauthoryear{{Ferguson}, {Dickinson}, \&
  {Papovich}}{{Ferguson} et~al.}{2002}]{Ferguson:02}
{Ferguson}, H.~C., {Dickinson}, M.,  \& {Papovich}, C. 2002, \apjl, 569, L65

\bibitem[\protect\citeauthoryear{{Giallongo} et~al.}{{Giallongo}
  et~al.}{2002}]{Giallongo:02}
{Giallongo}, E., {Cristiani}, S., {D'Odorico}, S.,  \& {Fontana}, A. 2002,
  \apjl, 568, L9

\bibitem[\protect\citeauthoryear{{Gnedin}}{{Gnedin}}{2000}]{Gnedin:00}
{Gnedin}, N.~Y. 2000, \apj, 535, 530

\bibitem[\protect\citeauthoryear{{Gnedin} \& {Ostriker}}{{Gnedin} \&
  {Ostriker}}{1997}]{GnedinO:97}
{Gnedin}, N.~Y.,  \& {Ostriker}, J.~P. 1997, \apj, 486, 581

\bibitem[\protect\citeauthoryear{{Harris} \& {Pudritz}}{{Harris} \&
  {Pudritz}}{1994}]{Harris:94}
{Harris}, W.~E.,  \& {Pudritz}, R.~E. 1994, \apj, 429, 177

\bibitem[\protect\citeauthoryear{{Hurwitz}, {Jelinsky}, \& {Dixon}}{{Hurwitz}
  et~al.}{1997}]{Hurwitz:97}
{Hurwitz}, M., {Jelinsky}, P.,  \& {Dixon}, W. V.~D. 1997, \apjl, 481, L31

\bibitem[\protect\citeauthoryear{{Kogut} et~al.}{{Kogut}
  et~al.}{2003}]{Kogut:03}
{Kogut}, A., et~al. 2003, \apjs, 148, 161

\bibitem[\protect\citeauthoryear{{Kundu} \& {Whitmore}}{{Kundu} \&
  {Whitmore}}{2001}]{KunduII:01}
{Kundu}, A.,  \& {Whitmore}, B.~C. 2001, \aj, 122, 1251

\bibitem[\protect\citeauthoryear{{Lanzetta} et~al.}{{Lanzetta}
  et~al.}{2002}]{Lanzetta:02}
{Lanzetta}, K.~M., {Yahata}, N., {Pascarelle}, S., {Chen}, H.,  \& {Fern{\'
  a}ndez-Soto}, A. 2002, \apj, 570, 492

\bibitem[\protect\citeauthoryear{{Leitherer} et~al.}{{Leitherer}
  et~al.}{1999}]{Leitherer:99}
{Leitherer}, C., et~al. 1999, \apjs, 123, 3

\bibitem[\protect\citeauthoryear{{McLaughlin}}{{McLaughlin}}{1999}]{McLaughlin%
:99}
{McLaughlin}, D.~E. 1999, \aj, 117, 2398

\bibitem[\protect\citeauthoryear{{McLaughlin} \& {Pudritz}}{{McLaughlin} \&
  {Pudritz}}{1996}]{McLaughlin:96}
{McLaughlin}, D.~E.,  \& {Pudritz}, R.~E. 1996, \apj, 457, 578

\bibitem[\protect\citeauthoryear{{Miralda-Escud\'e}, {Haehnelt}, \&
  {Rees}}{{Miralda-Escud\'e} et~al.}{2000}]{Miralda:00}
{Miralda-Escud\'e}, J., {Haehnelt}, M.,  \& {Rees}, M.~J. 2000, \apj, 530, 1

\bibitem[\protect\citeauthoryear{{Peebles} \& {Dicke}}{{Peebles} \&
  {Dicke}}{1968}]{Peebles:68}
{Peebles}, P. J.~E.,  \& {Dicke}, R.~H. 1968, \apj, 154, 891

\bibitem[\protect\citeauthoryear{{Persic} \& {Salucci}}{{Persic} \&
  {Salucci}}{1992}]{Persic:92}
{Persic}, M.,  \& {Salucci}, P. 1992, \mnras, 258, 14P

\bibitem[\protect\citeauthoryear{{Ricotti}}{{Ricotti}}{2002}]{Ricotti:02}
{Ricotti}, M. 2002, \mnras, 336, L33

\bibitem[\protect\citeauthoryear{{Ricotti}, {Gnedin}, \& {Shull}}{{Ricotti}
  et~al.}{2002a}]{RicottiGSa:02}
{Ricotti}, M., {Gnedin}, N.~Y.,  \& {Shull}, J.~M. 2002a, \apj, 575, 33

\bibitem[\protect\citeauthoryear{{Ricotti}, {Gnedin}, \& {Shull}}{{Ricotti}
  et~al.}{2002b}]{RicottiGSb:02}
{Ricotti}, M., {Gnedin}, N.~Y.,  \& {Shull}, J.~M. 2002b, \apj, 575, 49

\bibitem[\protect\citeauthoryear{{Ricotti} \& {Ostriker}}{{Ricotti} \&
  {Ostriker}}{2003a}]{RicottiOI:03}
{Ricotti}, M.,  \& {Ostriker}. 2003a, submitted, (astro-ph/0311003), (paper~I)

\bibitem[\protect\citeauthoryear{{Ricotti} \& {Ostriker}}{{Ricotti} \&
  {Ostriker}}{2003b}]{RicottiO:03}
{Ricotti}, M.,  \& {Ostriker}, J.~P. 2003b, submitted, (astro-ph/0310331),
  (paper~IIa)

\bibitem[\protect\citeauthoryear{{Ricotti} \& {Shull}}{{Ricotti} \&
  {Shull}}{2000}]{RicottiS:00}
{Ricotti}, M.,  \& {Shull}, J.~M. 2000, \apj, 542, 548

\bibitem[\protect\citeauthoryear{{Steidel}, {Pettini}, \&
  {Adelberger}}{{Steidel} et~al.}{2001}]{Steidel:01}
{Steidel}, C.~C., {Pettini}, M.,  \& {Adelberger}, K.~L. 2001, \apj, 546, 665

\bibitem[\protect\citeauthoryear{{Stetson}, {Vandenberg}, \& {Bolte}}{{Stetson}
  et~al.}{1996}]{Stetson:96}
{Stetson}, P.~B., {Vandenberg}, D.~A.,  \& {Bolte}, M. 1996, \pasp, 108, 560

\bibitem[\protect\citeauthoryear{{Taniguchi}, {Trentham}, \&
  {Ikeuchi}}{{Taniguchi} et~al.}{1999}]{Taniguchi:99}
{Taniguchi}, Y., {Trentham}, N.,  \& {Ikeuchi}, S. 1999, \apjl, 526, L13

\bibitem[\protect\citeauthoryear{{Wood} \& {Loeb}}{{Wood} \&
  {Loeb}}{2000}]{Wood:00}
{Wood}, K.,  \& {Loeb}, A. 2000, \apj, 545, 86

\end{thebibliography}

\label{lastpage}
\end{document}